\documentstyle[aps,prl,epsf,floats]{revtex}
\begin{document}
\twocolumn[{\hsize\textwidth\columnwidth\hsize\csname
@twocolumnfalse\endcsname
\title{
\draft
Ordering Periodic Spatial Structures by Noise
}
\author{
J. M. G. Vilar and J. M. Rub\'{\i}
}
\address{
Departament de F\'{\i}sica Fonamental, Facultat de
F\'{\i}sica, Universitat de Barcelona, Diagonal 647,
E-08028 Barcelona, Spain 
\date{\today}
}
\maketitle
\widetext
\begin{abstract}
\leftskip 54.8pt
\rightskip 54.8pt

We have analyzed the interplay between noise and periodic spatial
modulations in bistable systems outside equilibrium and
found that noise is able to increase
the spatial order of the
system, giving rise to periodic patterns which otherwise look random.
This new phenomenon, which may be viewed as the spatial counterpart
of stochastic resonance, then shows a constructive role of noise
in spatially extended systems, not considered up to now.

\end{abstract}
}]
\narrowtext
Contrarily to intuitive arguments, arising from the frequent identification
of noise with a source of disorder, 
in systems outside equilibrium the presence of noise
may be responsible for an increase of order.
Noise then can play a constructive role, losing its usually assumed character
of nuisance.
One of the most important manifestations of this constructive role played
by noise takes place when noise
and a weak signal periodic in time exhibit
a cooperative behavior, giving rise to the enhancement of the
periodic response.
This fact, which is evidenced by an increase of the output
signal-to-noise ratio (SNR) as the noise level increases, constitutes
the main fingerprint of the phenomenon known as  stochastic resonance (SR)
\cite{Benzi,Moss}.
The importance and interest of SR has been revealed
by the great number of situations in which it has been found
\cite{Maki,Maki2,JSP,Wies,Wiese,thre,BG,mio}.
In regards to spatially extended systems, this constructive role
of noise has only been reported
for situations in which also a time periodic signal enters
the system.
This feature occurring in the phenomenon known as
spatiotemporal stochastic resonance \cite{spi,array,phi4,dio,mio2}
only deals with
the effects of the spatial degrees of freedom in the
 time evolution 
of the system, but no enhancement of the spatial order,
in a similar fashion as the temporal order, has been found up to now.

In this  Letter we show 
that randomness may be responsible for spatial ordering
of a system in such a
way that the emergent  structures look more regular,
i. e. periodic, when the noise level
is increased.
This new phenomenon can be viewed as the spatial counterpart of SR.
In it, the SNR is defined through the structure
factor instead of the power spectrum and
the system is periodically modulated in space instead
of in time.
In this sense, noise is able to increase
the order of a spatial structure in a similar
fashion as noise is able to do it in the time evolution.

\begin{figure}[t]
\centerline{
\epsfxsize=8cm 
\epsffile{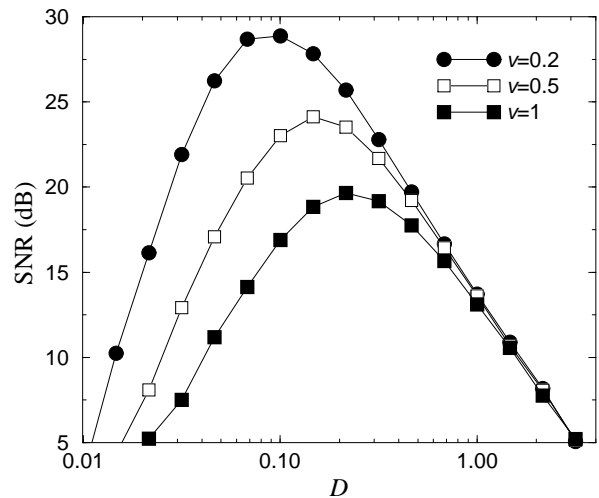}
}
\caption[b]{\label{figSNR}
SNR
as a function of the noise level
for different values of the advection velocity:
$v=0.2$, $0.5$, and $1$.
The values of the remaining parameters are
$r=1$, $g=1$, $\sigma=0.03$,
$A=0.3$, and  $k_0/2 \pi=0.05$.
}
\end{figure}

\begin{figure}[t]
\centerline{
\epsfxsize=8cm 
\epsffile{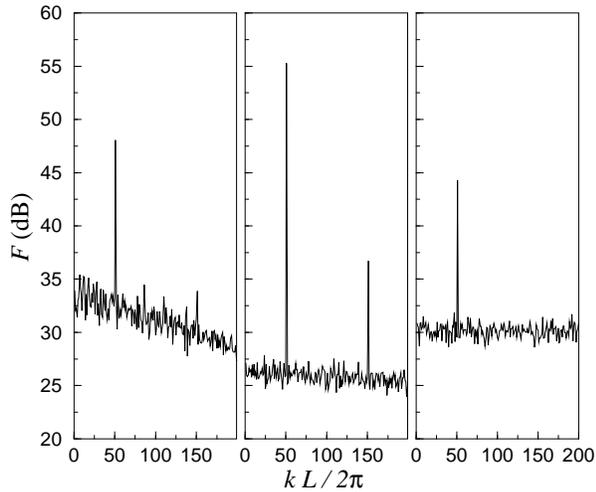}
}
\caption[b]{\label{figSF}
Structure factor
as a function of the dimensionless
wavenumber $kL/2 \pi$, with $L$ ($=1024$)  the
length of the system, for three  noise levels:
$D=0.02$, $0.1$, and $1$ (from left to right).
The values of the remaining parameters are
$v=0.2$,
$r=1$, $g=1$, $\sigma=0.03$,
$A=0.3$, and  $k_0L/2 \pi=50$.
}
\end{figure}

\begin{figure}[t]
\centerline{
\epsfxsize=8cm 
\epsffile{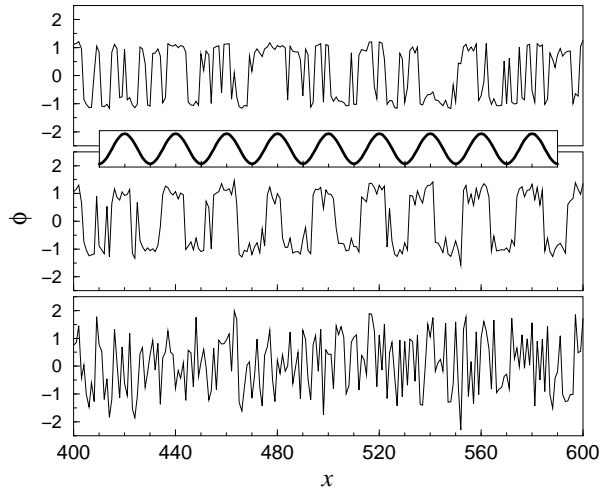}
}
\caption[b]{\label{figST}
Field $\phi$ as function of position $x$ for the same situation
as in Fig. \ref{figSF}
[$D=0.02$, $0.1$, and $1$ (from top to bottom)].
The inset illustrates the periodicity of the force
which accounts for the spatial modulation.
}
\end{figure}

\begin{figure}[t]
\centerline{
\epsfxsize=6.8cm 
\epsffile{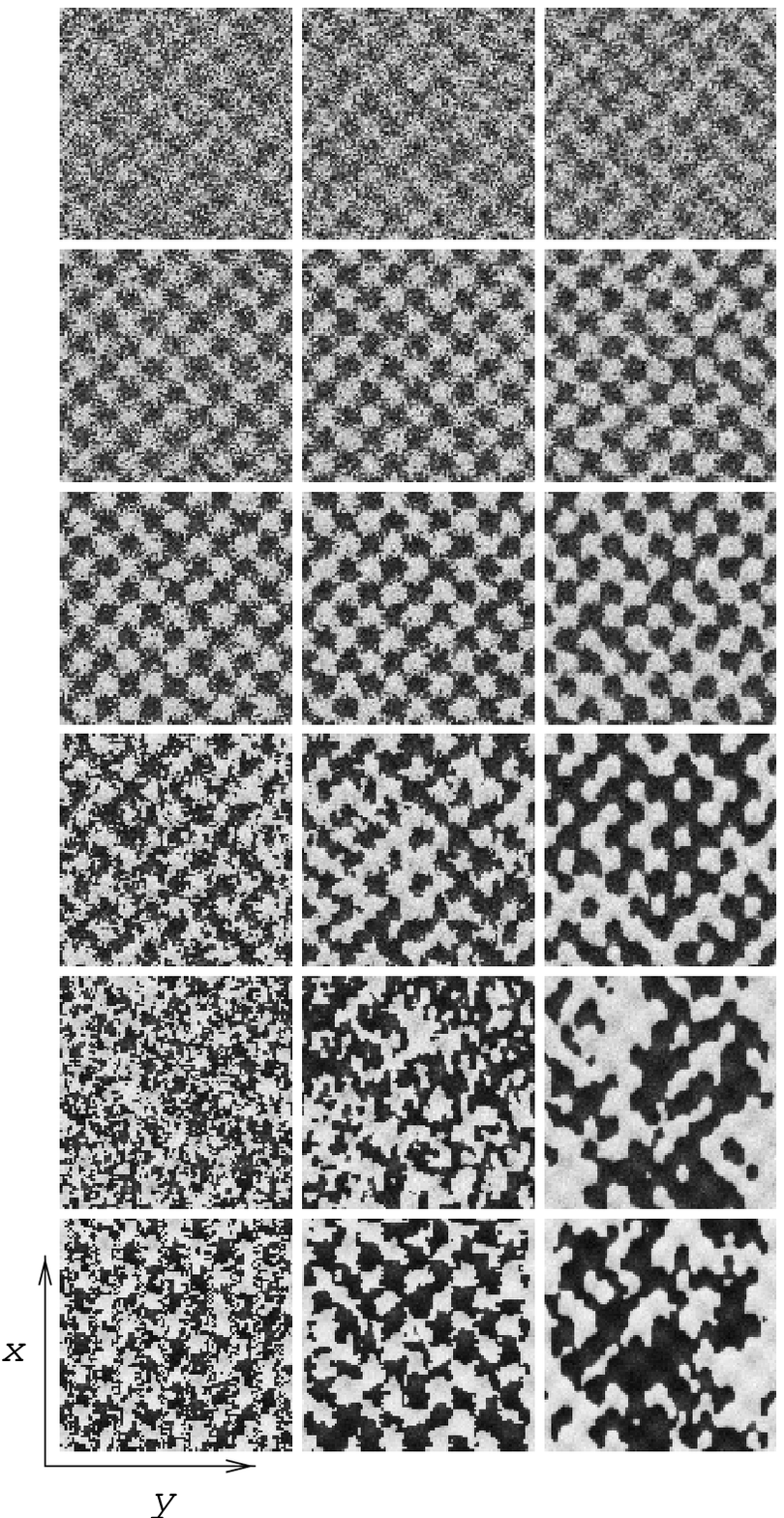}
}
\caption[b]{\label{figWN}
Representation of the two-dimensional field $\phi$ for
different values of the noise level
[$D=10^{-6/3}$, $10^{-5/3}$, $10^{-4/3}$,
$10^{-3/3}$, $10^{-2/3}$, and $10^{-1/3}$ (from bottom to top)]
and diffusion
[$\sigma=10^{-9/6}$, $10^{-7/6}$, and $10^{-5/6}$
(from left to right)].
The system size is $100 \times 100$.
The values of the remaining parameters are
$v=0.2$,
$r=1$, $g=1$,
$A=0.3$, and $k_0/2 \pi=0.05$.
Black and white colors stand for minimum and maximum values,
respectively.
}
\end{figure}

\begin{figure}[t]
\centerline{
\epsfxsize=6.8cm 
\epsffile{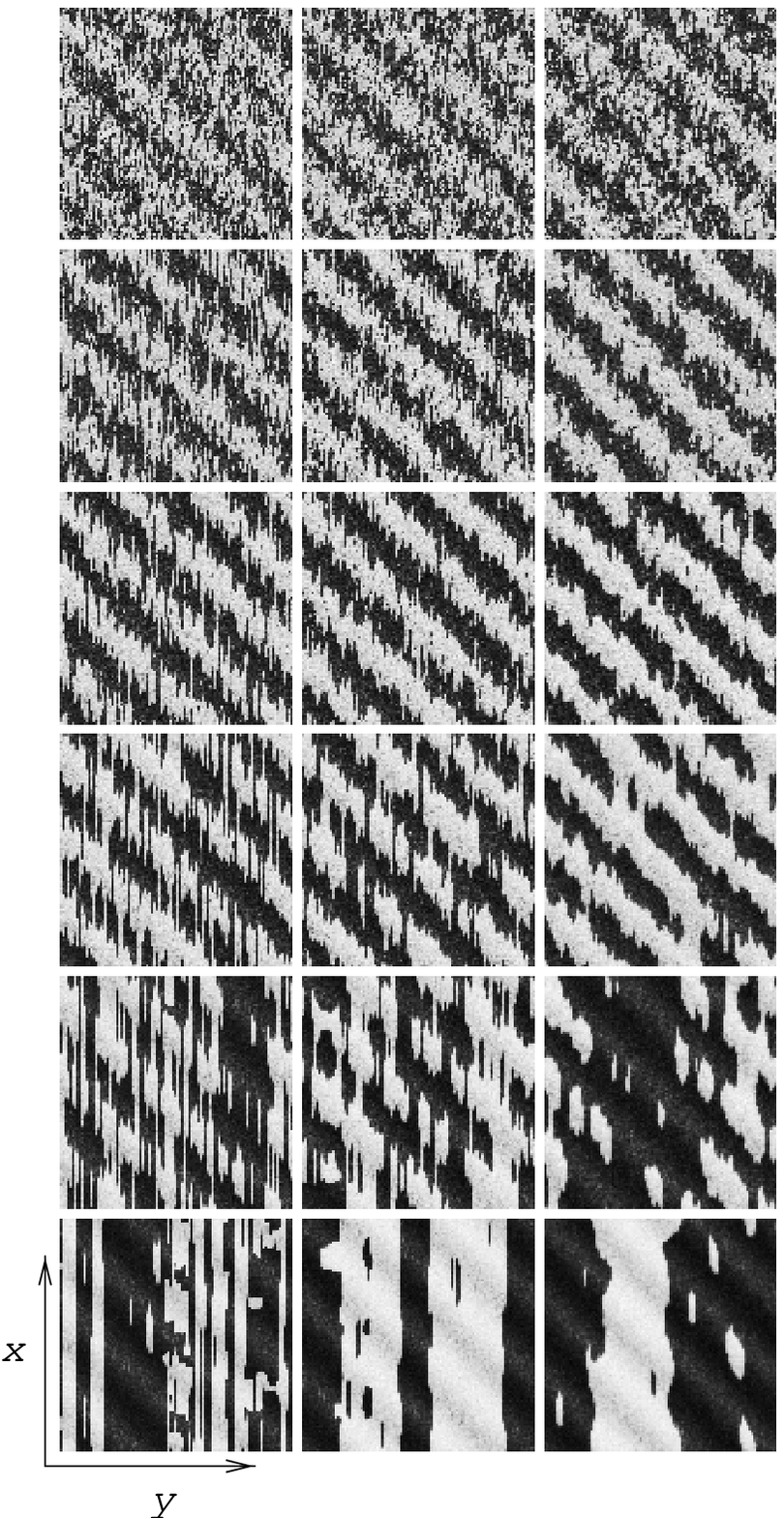}
}
\caption[b]{\label{figQN}
Representation of the two-dimensional field $\phi$ for
different values of the disorder intensity
[$D=10^{-6/3}$, $10^{-5/3}$, $10^{-4/3}$, $10^{-3/3}$,
$10^{-2/3}$, and $10^{-1/3}$ (from bottom to top)]
and diffusion
[$\sigma=10^{-9/6}$, $10^{-7/6}$, and $10^{-5/6}$ (from left to right)].
The system size is $100 \times 100$.
The values of the remaining parameters are
$v=0.5$,
$r=1$, $g=1$,
$A=0.3$, and $k_0/2 \pi=0.03$.
Black and white colors stand for minimum and maximum values,
respectively.
}
\end{figure}

To illustrate the essentials of the phenomenon we consider
explicitly  the $\phi^4$ model,
although similar results could also be obtained for other systems
provided that
they exhibit bistability.
Additionally, the system is under the influence of
a spatial periodic force and driven outside equilibrium  by advection.
We first study the one-dimensional case
in which the field $\phi$ is advected by a constant velocity $v$,
\begin{eqnarray}
\label{masterI}
\partial_t\phi + v\partial_x\phi & = &
r\phi -g\phi^3 + \sigma\partial_{xx}\phi \nonumber \\
& & +A\cos(k_0x)+\zeta(x,t) \;\; .
\end{eqnarray}
Here $A\cos(k_0x)$ is the periodic force,
accounting for the spatial modulation,
with $A$ the amplitude and
$2\pi/k_0$ the period. The parameters $r$, $g$ and $\sigma$
are positive constants, whereas
the noise term $\zeta(r,t)$ is Gaussian  and white  with zero mean and
correlation function
$\left<\zeta(x,t)\zeta(x^\prime,t^\prime)\right>
=2D\delta(x - x^\prime)\delta(t-t^\prime)$,
defining the noise level $D$.

With the purpose of characterizing the spatial configuration
we will consider the structure factor defined as
\begin{equation}
F(k)=\left\langle\hat\phi_k\hat\phi_{-k}\right\rangle \;\; ,
\end{equation}
where  $\hat\phi_k$ is the Fourier transform of the field
$\phi(x)$ and $\langle\;\rangle$ 
indicates average over the  noise  realization.
Due to the presence of the periodic force,
it consists of a delta
function centered at the wavenumber $k_0$ plus a function $Q(k)$
which is smooth in the neighborhood of $k_0$ and is given by
\begin{equation}
F(k)
=Q(k)+S(k_0)\delta(k-k_0) \;\;.
\end{equation}
The structure factor, as expressed previously,
explicitly shows the intensity of the periodic component
of the system, $S(k_0)$,  and the stochastic
component, $Q(k)$. The SNR, defined
as the ratio between both quantities,
\begin{equation}
\label{SNRdef}
\mbox{SNR}=S(k_0) / Q(k_0) \;,
\end{equation}
then indicates the order present in the
system.

To proceed further we have numerically integrated the
previous equations 
by discretizing  them on a mesh  \cite{press}
and then by using
standard methods for stochastic differential equations
\cite{kloeden}.
In Fig. \ref{figSNR} we have represented the SNR as a function
of the noise level for representative values of the parameters
corresponding to the bistable situation. 
This figure  indicates,
through the maximum of the SNR at a nonzero noise level,
that the presence of an optimum amount of
noise enhances the  underlying periodic structure of the system.
In this regard, in Fig. \ref{figSF}
we display the structure factor for the optimum,
a higher and a lower noise level.
The existence of a
periodic pattern is  revealed by the
peak arising over the background noise at $k_0$, which
for the optimum noise intensity, is more than
an order of magnitude higher than for the other displayed intensities.
An instance of how the enhancement of this peak, quantified by the
SNR,  manifest in the spatial structure is shown in Fig. \ref{figST}.
It is worth emphasizing the fact that the pattern for
the optimum noise intensity looks more deterministic, i. e. periodic,
than for lower noise intensities.
In this regard noise constitutes  a source of order.
Further increasing of the noise level, however,  destroys 
the coherent response to the periodic spatial modulation.

From Fig. \ref{figST} one can elucidate the mechanism
giving rise to the enhancement of
the structure.  For the previous values of the parameters
the system exhibits local bistability. The effect of the periodic
force is then to spatially modulate the system in such
a way that the most stable state changes from positive to
negative values of the field $\phi$, depending on position.
Since the field $\phi$ is advected, the system is unable to switch
between these two states when
noise level is too small.
The presence of an optimum amount of noise, however,
makes these transitions possible in a coherent fashion,
giving rise to the enhancement
of the underlying  pattern. Further increasing the
noise level completely destroys the coherent response.
In this context, this mechanism is the spatial counterpart
of the corresponding one of SR.
Thereby, the parameter $vk_0$, which has dimensions of the inverse
of time, plays the same role as the frequency $\omega_0$ in SR.
This fact is displayed in Fig. \ref{figSNR}, where the dependence
of the SNR on $v$, for fixed $k_0$, is close to that on $\omega_0$
in the case of SR.
The striking similarity with the temporal case is evidenced
even more when considering Fig. \ref{figST} for the 
optimum noise level, which looks quite similar  to
a bistable system in time when exhibiting SR.

To illustrate more the fact that noise
may imply ordering in nonequilibrium systems we have considered
the two-dimensional counterpart of Eq. (\ref{masterI}) on the $(x,y)$-plane.
The periodic force  is
now $A\cos(k_0x)\cos(k_0y)$, whereas the system is again
advected in the $x$-direction by a constant velocity $v$.
The other terms follow  by  the straightforward extension
to two dimensions. Hence, the system is described by
\begin{eqnarray}
\label{masterII}
\partial_t\phi + v\partial_x\phi & = &
r\phi -g\phi^3 + \sigma\left(\partial_{xx} +
\partial_{yy}\right) \phi \nonumber \\
& & +A\cos(k_0x)\cos(k_0y)+\zeta(x,y,t) \;\; .
\end{eqnarray}
In Fig. \ref{figWN} we have represented the field $\phi(x,y)$ obtained from
numerical simulations for  representative values of the
parameters.
For low and high noise level the spatial pattern is  completely disordered.
In contrast, an optimum nonzero amount of noise makes the 
presence of the underlying checkerboard pattern manifest. 
In this regard, the addition of noise is able to increase the order of
the system  giving rise to periodic structures which otherwise
would not be observed.
Notice that the periodicity in the perpendicular direction to the
advection is also enhanced.
It its worth emphasizing that, in contrast to the one-dimensional
case, in two dimensions a temporal counterpart does not exist
due to the one-dimensional character of time.

Another interesting situation comes when dealing with disorder
represented by quenched noise \cite{lind}. 
The noise term
is Gaussian with zero mean, but now with
correlation function
$\left<\xi(x,y)\xi(x^\prime,y^\prime)\right>
=2D\delta(x - x^\prime)\delta(y-y^\prime)$.
In this case, $D$ accounts  for the degree of disorder.
The explicit case we will study is
\begin{eqnarray}
\label{masterIII}
\partial_t\phi + v\partial_x\phi & = &
r\phi -g\phi^3 +
\sigma\left(\partial_{xx} + \partial_{yy}\right) \phi \nonumber \\
& & +A\cos[k_0(x+y)]+\zeta(x,y) \;\; ,
\end{eqnarray}
where, for the sake of generality, we have considered a
 new
expression for the periodic force.
The effects of disorder are illustrated in Fig. \ref{figQN}
which displays the field $\phi(x,y)$ for representative
values of the parameters.
In the figure one can see that for weak disorder,
i.e. low $D$, the pattern is quite
random, although the direction of the advection
is manifested in the elongated
form of the spots.
When increasing the disorder, the system is able to increase its order,
then displaying periodic striped patterns.
This constructive aspect, however, is lost
by further increasing the intensity of the quenched noise.
In this sense, there exists 
an optimum amount of disorder which is responsible for order.

It its worth pointing out that
the phenomenon we have found 
is even more general than previously presented,
since this kind of spatial ordering may
also appear when
advection is not present and even in systems without
intrinsic spatial structure.
For instance,
by only replacing  $x$ by $x-vt$ in Eq. (\ref{masterI})
the advective term disappears and the force $A\cos(k_0x)$
becomes $A\cos[k_0(x-vt)]$. Under these circumstances,
Eq. (\ref{masterI}) describes the propagation of a plane traveling wave
in a bistable medium. Thereby, the results we have obtained also apply
to this case.

In summary, we have found a new phenomenon in which
randomness is responsible
for the spatial ordering of a system.
In this regard, the addition 
of noise makes the presence of spatial periodicity manifest,
giving rise, for instance, to checkerboard and striped patterns.
We have shown the occurrence of the phenomenon
for different types of spatial modulations and different
types of randomness, such as white noise and spatial disorder.
It is worth emphasizing that although our analysis has
been explicitly carried out for the $\phi^4$ model,
similar results could also be obtained for other systems provided that
they exhibit bistability.
Our findings, then, contribute
to a wider understanding of the
role of nonequilibrium fluctuations in spatially extended systems,
indicating that this constructive
aspect of noise, reported before only for temporal signals, is
more universal than believed.

This work was supported by DGICYT of the Spanish Government under
Grant No.  PB95-0881.  J.M.G.V.  wishes to thank Generalitat de
Catalunya for financial support.


\begin{references}

\bibitem{Benzi} R. Benzi,  A.  Sutera, and A. Vulpiani,
J. Phys. A {\bf 14}, L453 (1981).

\bibitem{Moss}F. Moss, in
{\it Some Problems in Statistical Physics},
edited by G. H. Weiss (SIAM, Philadelphia,1994).

\bibitem{Maki}B. McNamara, K. Wiesenfeld, and R. Roy,
Phys. Rev. Lett. {\bf 60}, 2626-2629 (1988);

\bibitem{Maki2}B. McNamara and K. Wiesenfeld,
Phys. Rev. A {\bf 39}, 4854 (1989).

\bibitem{JSP}Proceedings of the NATO Advanced Research
Workshop on Stochastic Resonance, San Diego,
1992 [J. Stat. Phys. {\bf 70}, 1 (1993)]. 


\bibitem{Wies}K. Wiesenfeld,  D. Pierson, E. Pantazelou, 
C. Dames,  and F. Moss,
Phys. Rev. Lett. {\bf 72}, 2125 (1994).

\bibitem{Wiese} K. Wiesenfeld and  F. Moss,
Nature {\bf 373}, 33 (1995).

\bibitem{thre}Z. Gingl, L. B. Kiss, and F. Moss,
Europhys. Lett. {\bf 29} 191-196 (1995).

\bibitem{BG} A. Bulsara and L. Gammaitoni,
Phys. Today {\bf 49}, No. 3, 39 (1996).

\bibitem{mio} J. M. G. Vilar and J. M. Rub\'{\i},
Phys. Rev. Lett. {\bf 77}, 2863 (1996).


\bibitem{spi} P. Jung and G. Mayer-Kress,
Phys. Rev. Lett. {\bf 74}, 2130 (1995).

\bibitem{array} J. F. Lindner, B. K. Meadows, W. L. Ditto,
M. E. Inchiosa, and A. R. Bulsara,
Phys. Rev. Lett. {\bf 75}, 3 (1995).

\bibitem{phi4}F. Marchesoni, L. Gammaitoni,
and A. R. Bulsara,
Phys. Rev. Lett. {\bf 76}, 2609 (1996).

\bibitem{dio} M. L\"ocher, G. A. Johnson, and
E. R. Hunt,
Phys. Rev. Lett. {\bf 77}, 4698 (1996).

\bibitem{mio2} J. M. G. Vilar and J. M. Rub\'{\i},
Phys. Rev. Lett. {\bf 78}, 2886 (1997).

\bibitem{press} W.H. Press, B. P. Flannery,
S. A. Teukolsky,
and W. T. Vetterling, {\it Numerical Recipes} (Cambridge
University Press, New York, 1986).

\bibitem{kloeden} P. E. Kloeden and E. Platen,
{\it Numerical Solution of Stochastic
Differential Equations}
(Springer-Verlag, Berlin, 1995).

\bibitem{lind}
It has  been found that, paradoxically, the introduction of disorder
may force a chaotic system to exhibit a
periodic  behavior in time [Y. Braiman, J. Lindner, and W. Ditto,
Nature, {\bf 378}, 465 (1995)]. However, in such a situation the spatial
structure still remains disordered.


\end{references}
\end{document}